# Measurements of the Generalized Electric and Magnetic Polarizabilities of the Proton at Low $Q^2$ Using the VCS Reaction

P. Bourgeois,[1] Y. Sato,[2] J. Shaw,[1] R. Alarcon,[3] A. M. Bernstein,[4] W. Bertozzi,[4] T. Botto,[4] J. Calarco,[5] F. Casagrande,[4] M. O. Distler,[6] K. Dow,[4] M. Farkondeh,[4] S. Georgakopoulos,[7] S. Gilad,[4] R. Hicks,[1] M. Holtrop,[5] A. Hotta,[1] X. Jiang,[8] A. Karabarbounis,[7] J. Kirkpatrick,[5] S. Kowalski,[4] R. Milner,[4] R. Miskimen,[1] I. Nakagawa,[4,*] C. N. Papanicolas,[7] A. J. Sarty,[9] S. Sirca,[4] E. Six,[3] N. F. Sparveris,[7] S. Stave,[4] E. Stiliaris,[7] T. Tamae,[2] G. Tsentalovich,[4] C. Tschalaer,[4] W. Turchinetz,[4] Z.-L. Zhou,[4] and T. Zwart[4]

[1]*Department of Physics, University of Massachusetts, Amherst, Massachusetts 01003 USA*
[2]*Laboratory of Nuclear Science, Tohoku University, Mikamine, Taihaku, Sendai 982-0826 Japan*
[3]*Department of Physics and Astronomy, Arizona State University, Tempe, Arizona 85287, USA*
[4]*Department of Physics, Laboratory for Nuclear Science and Bates Linear Accelerator Center, Massachusetts Institute of Technology, Cambridge, Massachusetts 02139, USA*
[5]*Department of Physics, University of New Hampshire, Durham, New Hampshire 03824, USA*
[6]*Institute fur Kernphysik, Universitaet Mainz, Mainz, Germany*
[7]*Institute of Accelerating Systems and Applications and Department of Physics, University of Athens, Athens, Greece*
[8]*Department of Physics and Astronomy, Rutgers University, Piscataway, New Jersey 08854 USA*
[9]*Department of Astronomy and Physics, St. Mary's University, Halifax, Nova Scotia, Canada*

The mean square polarizability radii of the proton have been measured for the first time in a virtual Compton scattering experiment performed at the MIT-Bates out-of-plane scattering facility. Response functions and polarizabilities obtained from a dispersion analysis of the data at $Q^2=0.06$ GeV$^2$/c$^2$ are in agreement with $O(p^3)$ heavy baryon chiral perturbation theory. The data support the dominance of mesonic effects in the polarizabilities, and the increase of β with increasing $Q^2$ is evidence for the cancellation of long-range diamagnetism by short-range paramagnetism from the pion cloud.

PACS numbers: 13.60.Fz, 14.20.Dh, 13.40.Gp, 13.40.-f,

The electromagnetic polarizabilities of the nucleon provide a vital testing ground for theories of low-energy QCD and nucleon structure, and are of compelling experimental and theoretical interest.[1] In the case of atomic polarizabilities the electric polarizability is approximately equal to the atomic volume. By contrast, the electric polarizability of the nucleon is approximately $10^4$ times smaller than the nucleon volume, demonstrating the extreme stiffness of the nucleon relative to the atom. Although the electric and magnetic polarizabilities of the proton, α and β, are known with reasonable accuracy[2] from real Compton scattering (RCS), much less is known about the polarizability distributions inside the nucleon. To measure these distributions it is necessary to use the virtual Compton scattering (VCS) reaction[3], where the incident photon is virtual. At low $Q^2$ it is expected[4] that $α(Q^2)$ should decrease with increasing $Q^2$ with a characteristic length scale given by the pion range. The first VCS experiments at Mainz[5] at $Q^2=0.33$ GeV$^2$/c$^2$ and later at JLab[6] at $Q^2=0.92$ and $1.76$ GeV$^2$/c$^2$ established that $α(Q^2)$ is falling off, but with a form inconsistent with a simple dipole shape[6]. By contrast, because of the cancellation of short-range paramagnetism with longer-range diamagnetism from the pion cloud, $β(Q^2)$ is predicted to grow with increasing $Q^2$ and peak[4,7] near $Q^2=0.1$ GeV$^2$/c$^2$. This letter reports on a VCS experiment on the proton performed at the out-of-plane scattering facility at the MIT-Bates linear accelerator at $Q^2=0.06$ GeV$^2$/c$^2$. Data taken at this low $Q^2$ can provide a test of chiral perturbation theory (ChPT), and are sensitive to the mean square electric and magnetic polarizability radii.

The relationship between VCS cross sections and the polarizabilities is most easily seen in the low energy expansion (LEX) of the unpolarized VCS cross section[3],

$$d^5\sigma^{VCS} = d^5\sigma^{BH+Born} + q'\Phi\Psi_0(q,\varepsilon,\theta,\varphi) + O(q'^2) \quad (1)$$

where $q(q')$ is the incident (final) photon 3-momenta in the photon-nucleon C.M. frame, $\varepsilon$ is the photon polarization, $\theta(\phi)$ is the C.M. polar (azimuthal) angle for the outgoing photon, and $\Phi$ is a phase space factor. $d^5\sigma^{BH+Born}$ is the cross section for the Bethe-Heitler + Born amplitudes only, i.e. no nucleon structure, and is exactly calculable from QED and the nucleon form factors. The polarizabilites enter the cross section expansion at order $O(q')$ through the term[8]

$$\Psi_0 = V_1\left[P_{LL}(q) - \frac{P_{TT}(q)}{\varepsilon}\right] + V_2 P_{LT}(q) \quad (2)$$

where $P_{LL}$, $P_{TT}$ and $P_{LT}$ are VCS response functions, with $P_{LL} \propto \alpha(Q^2)$, $P_{LT} \propto \beta(Q^2)$ +spin-polarizabilities, and $P_{TT} \propto$ spin-polarizabilities. $V_1$ and $V_2$ are kinematic functions. The Bates VCS experiment was designed to make an azimuthal separation of $P_{LL} - P_{TT}/\varepsilon$ and $P_{LT}$ by taking data simultaneously at $\phi$ angles of 90°, 180° and 270°, at fixed $\theta = 90°$. For the out-of-plane cross sections at $\phi = 90°$ and 270°, the cross sections are equal and the polarizability effect is proportional to $P_{LL} - P_{TT}/\varepsilon$. At $\phi = 180°$, the cross section is proportional to the sum of $P_{LL} - P_{TT}/\varepsilon$ and $P_{LT}$. Data were taken at five different C.M. final photon energies $q'$ ranging from 43 MeV, where the polarizability effect is negligible, up to 115 MeV where the polarizability effect is approximately 20%. The data were taken at q=240 MeV, and $\varepsilon = 0.9$, corresponding to $Q^2 \approx 0.06$ GeV$^2$/c$^2$.

The experiment was the first to use extracted CW beam from the MIT-Bates South Hall Ring. The extracted beams had duty factors of approximately 50%, currents of up to 7 $\mu A$, and the five beam energies ranged from 570 to 670 MeV. The target was 1.6 cm of liquid hydrogen. The experiment marked the first use of the full Out-of-Plane Spectrometer (OOPS) system with gantry for proton detection,[9] and a new OHIPS electron spectrometer focal plane[10] that increased the momentum acceptance of the spectrometer from 9% to 13%, giving increased acceptance in $q'$. Optics studies were performed to measure OHIPS transport matrix elements over the extended focal plane instrumentation, and a new OOPS optics tune using a 2.5 m drift distance was developed for the running at $q' = 43$ and 65 MeV because of the close packing of the OOPS's at those energies. Data taken at higher $q'$ used the standard 1.4 m drift for the OOPS. The lowest proton kinetic energy in the experiment was 30 MeV, and the OOPS trigger was modified to a two-fold trigger of the first two scintillators in the focal plane to increase trigger efficiency. A GEANT simulation of the OOPS trigger predicts a trigger efficiency of $\approx 99\%$. The acceptance montecarlo was based on the program Turtle[11], and measured spectrometer matrix elements were used for calculating focal-plane coordinates from target coordinates. The multiple scattering model[12] from GEANT4 was implemented in the acceptance montecarlo. Good agreement was achieved between measured and calculated angular and momentum distributions.

The final state photon was identified through missing mass and time-of-flight techniques. Photon yields were obtained by fitting the missing mass squared (MM$^2$) distributions using the radiated line shape calculated with the montecarlo and an empirical background to account for $A(e,e'p)X$ events on the havar target cell wall. Polynomial and skewed gaussian shapes for the MM$^2$ backgrounds gave identical yields within errors to fits that used the accidental $MM^2$ distributions for the background shape, and the latter distribution was utilized for peak fitting. Radiative corrections were applied to the data,[13] approximately 22% in these kinematics.

The VCS cross sections are shown in Fig. 1 with the statistical and systematic errors combined in quadrature. The dominant error is statistical, with the largest systematic uncertainty the OOPS tracking efficiency, $\approx 1.6\%$. Tests of the data normalization by elastic p(e,e′p) measurements were limited by leakage current from the Faraday cup. However, at the $200\times$ higher beam current of the VCS production runs versus the elastic runs, the uncertainty in the beam charge resulting from the leakage current is relatively small, approximately 0.8% at q′ =100 MeV and $\approx 0.2\%$ at the other q′ settings.

The solid lines in Fig. 1 are the Bethe-Heitler+Born (BH+Born) calculations, i.e. no polarizability effect, using Hoehler form factors.[14] The agreement between data and the BH+Born calculation is good at low q′, while at higher q′ the out-of-plane data falls significantly below the calculation because of destructive interference between the BH+Born and polarizability amplitudes. The in-plane cross sections show a much smaller deviation from the BH+Born cross sections at high q′ because the kinematic multipliers $V_1$ and $V_2$ in Eq. (2) have the same sign, and therefore much of the polarizability effect is canceled at O(q′). The dashed lines in Fig. 1 are fits to the data using the LEX, giving $P_{LL} - P_{TT}/\varepsilon = 54.5 \pm 4.8 \pm 2.0$ GeV$^{-2}$, and $P_{LT} = -20.4 \pm 2.9 \pm 0.8$ GeV$^{-2}$, where the first error is statistical and the second is systematic. The largest systematic error results from the ±0.1% uncertainty in the beam energies, which introduces an error in the response functions through the energy dependence of $d^5\sigma^{BH+Born}$. A LEX analysis using the Friedrich-Walcher form factors[15] gives identical results, within errors, to the analysis presented here using the Hoehler form factors. The LEX result for $P_{LL} - P_{TT}/\varepsilon$ is shown in Fig. 2, where the statistical and systematic errors have been combined in quadrature. Also shown in the figure is the parameter free O(p$^3$) calculation in heavy baryon chiral perturbation theory (HBChPT)[4], which is in good agreement with experiment for $P_{LL} - P_{TT}/\varepsilon$. However, the LEX result for $P_{LT}$ (not shown in Fig. 2) is much larger than the the RCS result and the HBChPT prediction.

A dispersion analysis of the data was performed using the VCS dispersion model[16]. In this analysis the VCS amplitudes are obtained from the MAID $\gamma^*p \to N\pi$ multipoles,[17] and the unconstrained asymptotic contributions to 2 out of the 12 VCS amplitudes are varied to fit the VCS data. The dotted curves in Fig. 1 show the best dispersion fits to the VCS cross sections. The polarizabilities are found by summing the fitted asymptotic terms with calculated $\pi N$ dispersive contributions, and $P_{LL} - P_{TT}/\varepsilon$ and $P_{LT}$ are obtained from $\alpha$, $\beta$ and spin-polarizabilities calculated in the dispersion model. The best fit response functions from the dispersion analysis are $P_{LL} - P_{TT}/\varepsilon = 46.7 \pm 4.9 \pm 2.0$ GeV$^{-2}$ and $P_{LT} - -8.9 \pm 4.2 \pm 0.8$ GeV$^{-2}$. The dispersion results are shown in Fig. 2 with the statistical and systematic errors combined in quadrature. The dispersion result for $P_{LL} - P_{TT}/\varepsilon$ is in near agreement with the LEX analysis and the HBChPT predictions. The dispersion result for $P_{LT}$ is in good agreement with the HBChPT prediction, and is much smaller than the LEX result.

The source of disagreement between the LEX and dispersion analyses for $P_{LT}$ is the near cancellation of the electric and magnetic polarizability responses at O(q′) for the in-plane kinematics, causing the polarizability effect to be predominately quadratic in q′. The LEX analysis is only valid in kinematics where the polarizability effect is linear in q′ (see Eq. (1)), while the dispersion analysis is valid to all orders in q′.[18]

The dispersion model fits give $\alpha = 7.85 \pm 0.87 \pm 0.36 \times 10^{-4}$ fm$^3$, and $\beta = 2.69 \pm 1.48 \pm 0.28 \times 10^{-4}$ fm$^3$., and these results are shown in Fig. 3 with the statistical and systematic errors combined in quadrature, along with previous results from RCS[2], Mainz[19] and JLab[6]. The Bates results for $\alpha$ and $\beta$ are in near agreement with the HBChPT prediction, shown as the solid curves in Fig. 3. The Bates data supports an increase of $\beta(Q^2)$ from the real photon point with a confidence level of 82%. In ChPT this increase results from the cancellation of long-range diamagnetism by short-range paramagnetism, both from the pion cloud.[4,7] Paramagnetism from the $\Delta(1232)$ is predicted to be nearly independent of $Q^2$ in this low $Q^2$ range.[4]

The mean square electric polarizability radius $<r_\alpha^2>$ was determined from a dipole fit to the RCS and Bates $\alpha(Q^2)$ data points, giving $<r_\alpha^2> = 1.95 \pm 0.33$ fm$^2$, which is in agreement with the HBChPT prediction[20] of 1.7 fm$^2$. The experimental value is significantly larger than the proton mean square charge radius[21] of $0.757 \pm .014$ fm$^2$, which is evidence for the dominance of mesonic effects in the electric polarizability. It is interesting to note that the experimental result is close to the uncertainty principle estimate of 2.0 fm$^2$ for the size of the pion cloud.

An O($p^4$) calculation[22] of $\beta(Q^2)$ shows a nearly linear increase with $Q^2$ in the low $Q^2$ region, and therefore a straight line fit to the RCS and Bates $\beta(Q^2)$ data points was used to make an estimate of the sign and size of $<r_\beta^2>$. The value obtained from this fit, $<r_\beta^2> = -1.91 \pm 2.12$ fm$^2$, is in good agreement with the HBChPT prediction[20] of $-2.4$ fm$^2$.

The experimental results from this experiment are summarized in table 1. The experiment supports two long accepted, although arguably not fully tested, tenets of proton polarizabilities. The first is that the electric polarizability is dominated by mesonic effects, and this is confirmed by the size of $<r_\alpha^2>$. The second is the cancellation of positive short-range paramagnetism by longer-ranged diamagnetism, often envoked to explain the small size of $\beta$ relative to $\alpha$, and evidence for this is seen in the increase of $\beta$ with increasing $Q^2$. In the near future data on the spin-polarizabilities of the nucleon will be forthcoming from experiments at Mainz[23] and TUNL/HIGS[24] in experiments utilizing polarized beam, targets and recoil polarimetry.

The authors thank T. Hemmert, B. Holstein, I. L'vov, B. Pasquini, and M. Vanderhaeghen for their comments and for communicating the results of their calculations. The authors also thank the staff of the MIT-Bates linear accelerator facility for their efforts on this experiment. This work was supported in part by the program Pithagoras of the Hellenic Ministry of Education, by Grant-in-Aid for Scientific Research from the Japan Society for the Promotion of Science (KAKENHI, grants 14540239 and 17540229), and by D.O.E. grant DE-FG02-88ER40415.

Table 1. Response function units are GeV$^{-2}$, the polarizabilities $10^{-4}$ fm$^3$, and the mean square radius fm$^2$. The errors are statistical and systematic, respectively.

|  | LEX analysis | Dispersion analysis |
|---|---|---|
| $P_{LL} - P_{TT}/\varepsilon$ | $54.5 \pm 4.8 \pm 2.0$ | $46.7 \pm 4.9 \pm 2.0$ |
| $P_{LT}$ | - | $-8.9 \pm 4.2 \pm 0.8$ |
| $\alpha(Q^2 = 0.06)$ | - | $7.85 \pm 0.87 \pm 0.36$ |
| $\beta(Q^2 = 0.06)$ | - | $2.69 \pm 1.48 \pm 0.28$ |
| $<r_\alpha^2>$ | - | $1.95 \pm 0.33$ |

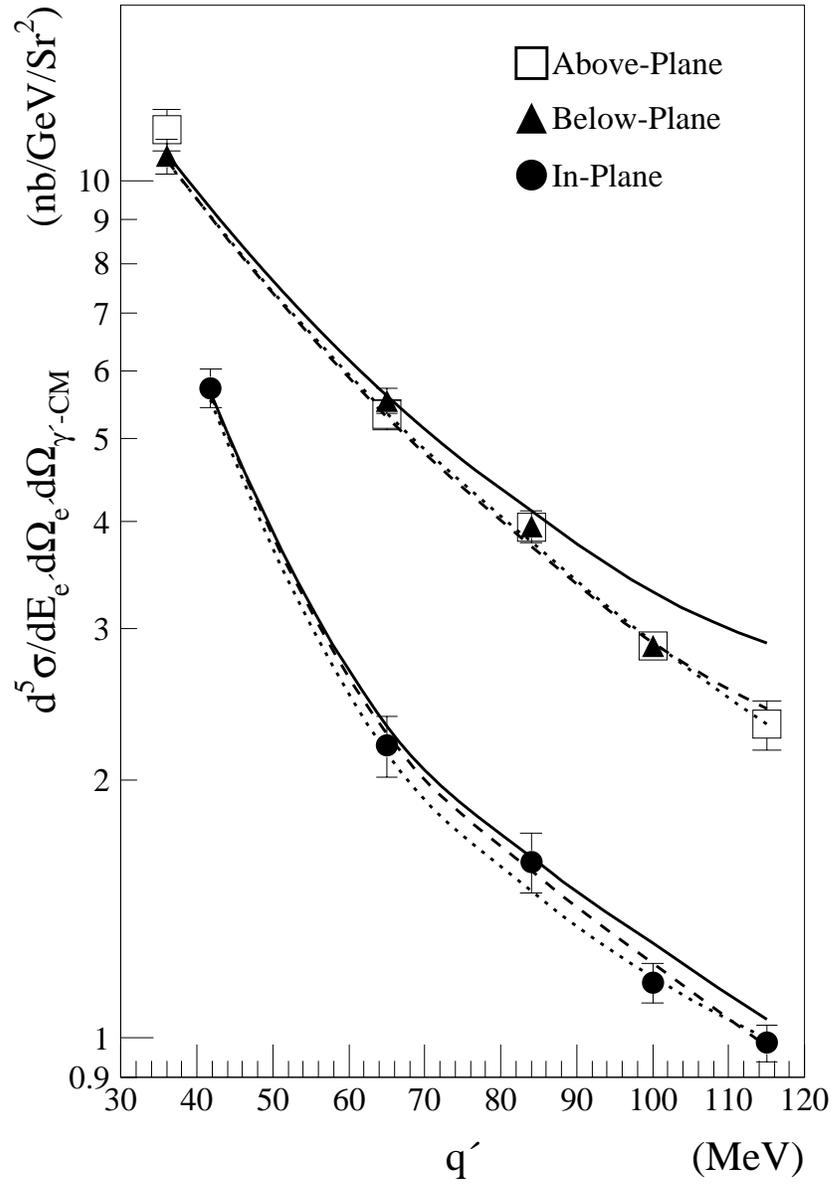

Fig.1. VCS cross sections as a function of $<q'>$. The solid curves are Bethe-Heitler+Born, the dashed and dotted curves are fits with LEX and dispersion analyses, respectively.

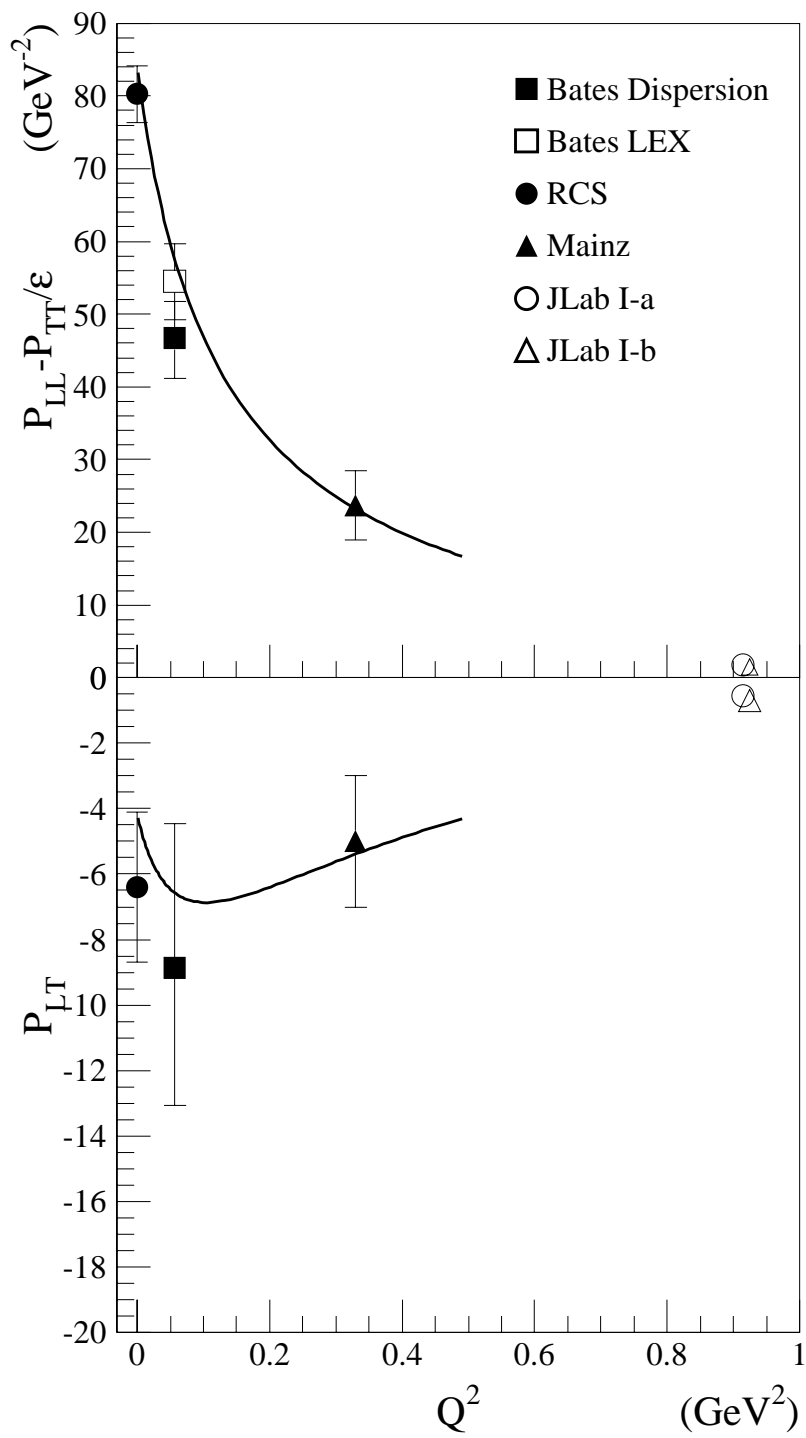

Fig. 2. VCS response functions from this experiment, RCS[2], Mainz[5] and JLab[6]. The solid curves are $O(p^3)$ HBChPT[4].

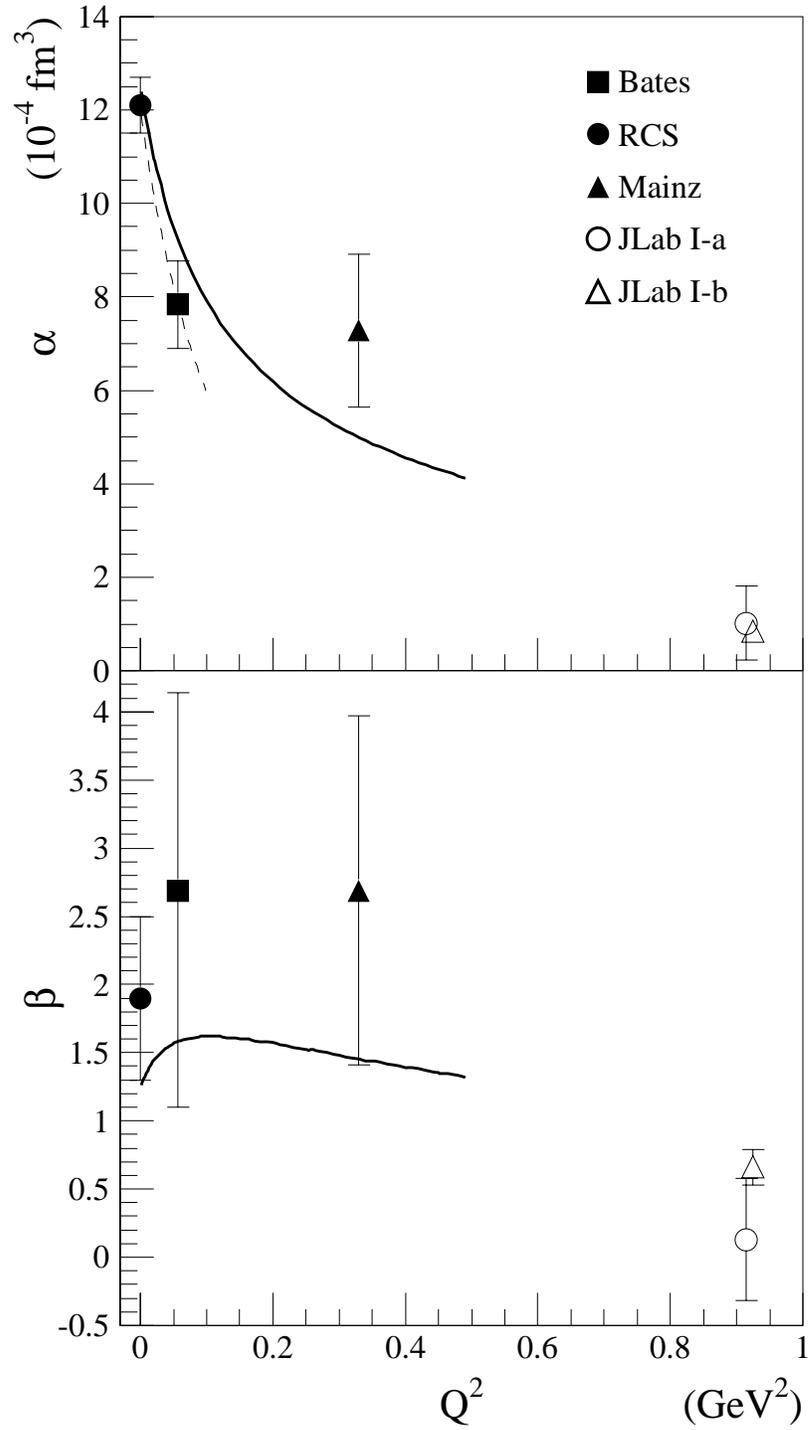

Fig.3. Dispersion analysis results for $\alpha(Q^2)$ and $\beta(Q^2)$. The references are the same as in Fig. 2 except for Mainz[19]. The solid curves are $O(p^3)$ HBChPT[4], the dashed curve is a low $Q^2$ fit to $\alpha(Q^2)$.